# Compact Reconnaissance Imaging Spectrometer for Mars (CRISM) north polar springtime recession mapping: First Three Mars years of observations

Adrian J. Brown*[1], Wendy M. Calvin[2], Scott L. Murchie[3]

[1] *SETI Institute, 189 Bernardo Ave, Mountain View, CA 94043, USA*
[2] *Geological Sciences, University of Nevada, Reno, NV, 89557, USA*
[3] *Johns Hopkins University Applied Physics Laboratory, Laurel, MD, 20723, USA*

Corresponding author:
Adrian Brown
SETI Institute
189 Bernardo Ave, Mountain View, CA 94043
ph. 650 810 0223
fax. 650 968 5830
email. abrown@seti.org

## ABSTRACT

We report on mapping of the north polar region of Mars using data from the Compact Reconnaissance Imaging Spectrometer for Mars (CRISM) instrument. We have observed three Mars Years (28-30) of late-winter and spring recessions ($L_s$=304°-92°). Our investigations have led to the following observations:

1. We classify the retreat of the north polar seasonal cap into 'pre-sublimation', 'early spring', 'asymmetric' and 'stable' periods according to the prevalent $H_2O$ ice grain size distributions.
2. During the early spring, the signatures of $CO_2$ ice at the edge of the cap are obscured by $H_2O$ ice, which increases the apparent size of the $H_2O$ ice annulus around the seasonal $CO_2$ cap at this time. At around $L_s$=25°, this process changes into an asymmetrical distribution of $H_2O$ deposition, covering $CO_2$ signatures more rapidly in the longitude range from 90-210°E.
3. We detect signatures of 'pure' $CO_2$ ice in extremely limited locations (in Lomonosov Crater) even in mid winter. $H_2O$ ice signatures appear everywhere in the retreating $CO_2$ seasonal cap, in contrast with the south polar seasonal cap.
4. We find that average $H_2O$ ice grain sizes continuously increase from northern mid-winter to the end of springtime – this is the inverse of the behavior of $CO_2$ ice grain sizes in the southern springtime.

## KEYWORDS

Mars, Polar regions, NPLD, $CO_2$ ice, $H_2O$ ice, grain size

* corresponding author, email: abrown@seti.org

## INTRODUCTION

The Martian polar regions are key to understanding the current climate and energy balance of the Red Planet. More than 25% of the Martian atmosphere participates in the polar sublimation and condensation cycle [*Hess, et al.*, 1977; *Hess, et al.*, 1979; *Hess, et al.*, 1980; *Kelly, et al.*, 2006]. This dynamic movement of carbon dioxide between the atmosphere and the surface is important to understand since it has implications for past, present and future Martian habitability. Even as we study this process in greater detail, we now have information at hand that suggests the cycle itself may be changing [*Haberle and Kahre*, 2010] and that reservoirs of $CO_2$ ice are sequestered in the polar regions and not currently interacting with the Martian seasonal cycle [*Phillips, et al.*, 2011].

This study reports the seasonal changes in the north polar region of Mars as observed by the Compact Imaging Spectrometer for Mars (CRISM) and the Mars Color Imager (MARCI) camera data on the Mars Reconnaissance Orbiter (MRO) spacecraft during its first three Mars years in orbit. Mars Years are abbreviated MY and MY1 started 11 April 1955 [*Clancy, et al.*, 2000] - here we report on observations from MY 28-30. An accompanying study of the south polar region was recently reported in *Brown et al.* [2009] and a similar study has been completed on a smaller scale at Louth Crater [*Brown, et al.*, 2008a].

Northern springtime recession have been observed in the infrared using orbiting instruments by *Kieffer and Titus* [2001] using the Thermal Emisson Spectrometer (TES) and *Appere et al.* [2011] using the OMEGA VNIR spectrometer. Spring recession visible albedos have been tracked by spacecraft using Viking cameras ([*James*, 1979; 1982]), Mars Orbiting Camera (*Bass et al.* [2000], *James and Cantor* [2001], *Benson and James*, [2005]), Mars Orbiting Laser Altimeter (*Byrne et al.* [2008]) and MARCI (*Cantor et al.* [2011]). The retreating $CO_2$ north polar cap is surrounded by a $H_2O$ ice annulus that is deposited by $H_2O$ ice carried onto the cap by baroclinic eddies which is cold-trapped on top of the $CO_2$ ice cap in what has been termed the 'Houben process' [*Houben, et al.*, 1997; *Bass and Paige*, 2000; *Schmitt, et al.*, 2006; *Wagstaff, et al.*, 2008; *Appere, et al.*, 2011]. *Appere et al.* discovered that this process was not symmetrical. We show here new maps of this asymmetry that lead us to the conclusion that the Houben process (where $H_2O$ ice is cold trapped on the edge of the retreating $CO_2$ ice cap by on-cap winds [*Houben, et al.*, 1997]) is not symmetrical around the cap and from $L_s$=25°-62° water ice is deposited on and obscures more than half of the remaining $CO_2$ ice cap. We present a new model as a possible explanation of this intriguing asymmetrical $H_2O$ ice deposition.

# METHODS

## CRISM and MARCI

The CRISM instrument is a visible to near infrared imaging spectrometer with spectral coverage of the ~0.36-3.92 μm range [*Murchie, et al.*, 2007]. The S-channel detector on CRISM covers the 0.362-1.053 μm range and the L-channel from 1.002 to 3.92 μm. We used only L-channel spectra in this study.

CRISM operates with a gimbal to obtain full and half resolution images suitable for geological mapping at ~18m/pixel. In addition, CRISM can operate in a nadir pointing 'mapping mode' to collect 10x binned pixels (~182.5m/pixel) with a smaller number of bands for reduced data rate for global mapping. The observations reported here are all taken in CRISM mapping mode. The CRISM swath width on the ground is a narrow ~10km, which is responsible for the 'spaghetti strand' appearance of maps in the polar region (Figure 1).

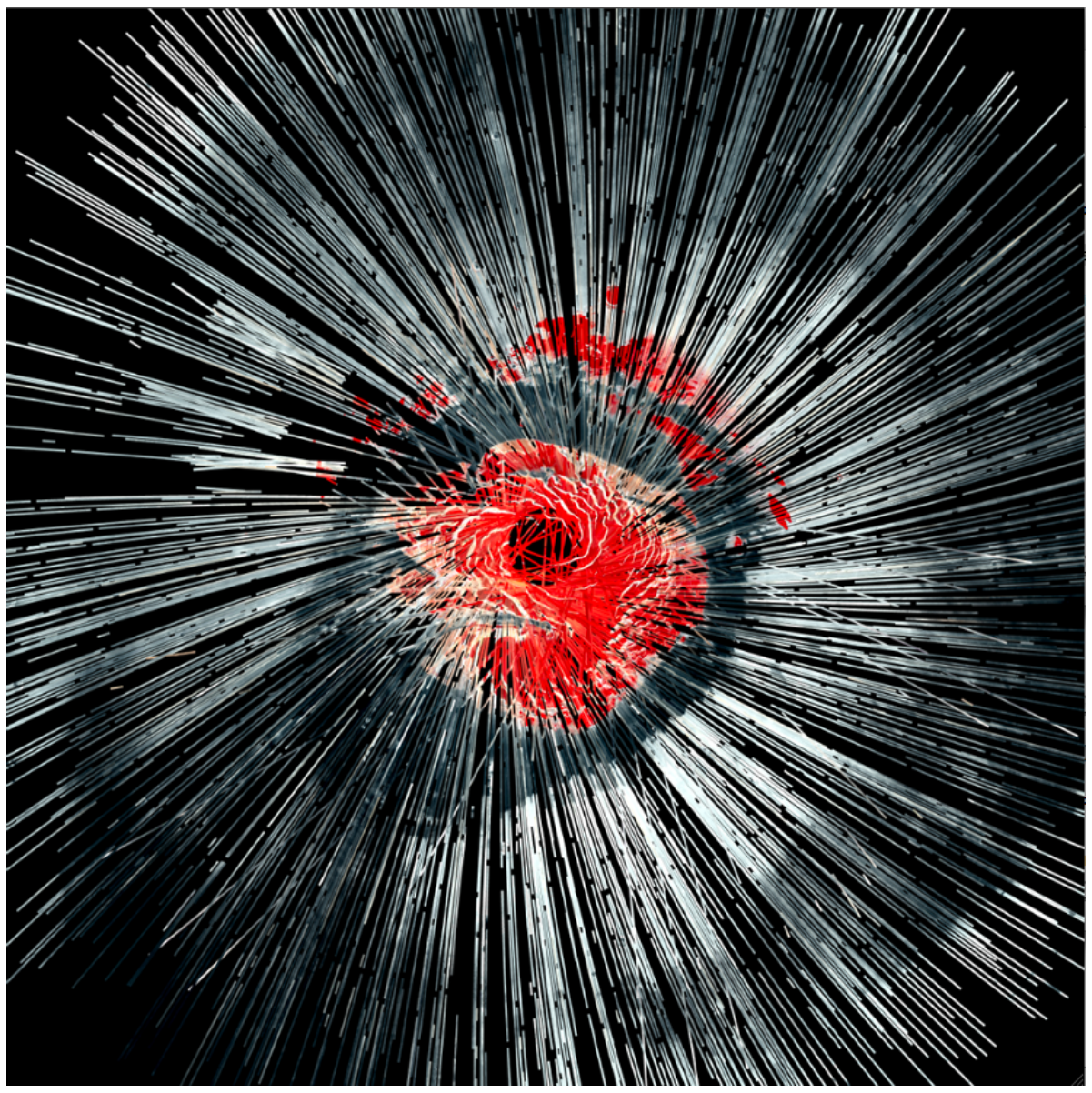

Figure 1. North polar mosaic of summertime CRISM MSP images from $L_s$=128°-160°. Red shading indicates the strength of the $H_2O$ ice index (Eqn 1).

The MARCI instrument is a wide angle 180 field of view camera that images Mars every day with 12 continuous limb to limb color images, each covering 60 of longitude. MARCI has 2 UV channels and 5 visible channels. The Visible channels used in this paper have 1km resolution [*Malin, et al.*, 2001].

## CRISM Seasonal mosaics

For this study, we have collected all mapping observations acquired north of 55° latitude and constructed seasonal mosaics of observations based on the two week MRO planning sequence using the MR PRISM software suite [*Brown and Storrie-Lombardi*, 2006]. All maps are presented in a polar stereographic projection from 55°N to the pole (Figure 1). Mosaics were constructed pixel by pixel, and overlapping data was overwritten by the last image acquired during the period. No averaging of pixels was attempted.

CRISM pixels were processed with a '*cos(i)*' correction using the incidence angles supplied with the CRISM image backplane data based on MOLA topographical data. No atmospheric correction has been attempted. All albedos discussed in this paper are therefore apparent Lambert albedos, which assume isotropic scattering of incident light back to the observer. More information on CRISM processing is presented in *Brown et al.* [2009].

CRISM collects mapping data in two forms - CRISM MSP (MultiSpectral Polar, which is a multispectral mapping mode) and HSP (Hyperspectral Polar) images. These are collected in nadir mode and have a reduced set of bands to minimize data rates. MSP observations have 19 'S' (short wavelength) bands and 55 'L' channel bands, HSP observations have 107 'S' channel bands and 154 'L' channel bands.

CRISM commenced imaging of the north polar region in September 2006, in the 2 week transition phase prior to commencement of the MRO science mission phase. MRO suffered unexplained resets of the onboard command computer during 2009 and was placed in maintenance mode from 26 August-16 December 2009 (late northern winter MY29 to early northern spring MY30), during which time no CRISM data was acquired. During September 2006-January 2010, only MSP-type mapping observations were collected. In MY30 (13 January 2010) HSP-type mapping observations were added in response to unanticipated extra MRO-downlink capability. After 13 January 2010, the CRISM instrument collected HSP and/or MSP mapping data as determined by the JHU/APL science team based on variable data downlink capacity.

A summary of north polar mapping observations for each MRO planning period is presented in Table 1.

| MRO Planning Cycle | DOY (2006-10) | (MY) and $L_s$ Range (°) | S channel observations | L channel observations |
|---|---|---|---|---|
| **1(TRA)** | **270-276** | **(28) 112.2-115.0** | **59** | **69** |
| **2** | **312-325** | **132.1-138.4** | **412** | **416** |
| **3** | **326-339** | **138.9-145.4** | **510** | **520** |
| **4** | **340-353** | **145.9-152.5** | **325** | **333** |
| **5** | **354-(07)002** | **153.1-159.9** | **534** | **535** |
| **6** | **003-016** | **160.4-167.3** | **343** | **346** |
| **7** | **017-029** | **167.9-174.5** | **140** | **162** |
| **8** | **031-044** | **175.6-182.9** | **114** | **138** |
| **9** | **046-058** | **184.0-190.9** | **126** | **126** |
| 10 | 070-072 | 198.0-199.2 | 13 | 13 |
| 11 | 082-083 | 205.2-205.8 | 7 | 7 |
| 14 | 115 | 225.4 | 1 | 1 |
| No north observations during northern fall (24 Mar-5 Aug 2007) | | | | |
| 21 | 217-219 | 289.8-291.2 | 10 | 11 |
| 22 | 227-240 | 295.7-304.0 | 27 | 27 |
| 23 | 241-253 | 304.1-311.6 | 72 | 72 |
| 24 | 255-268 | 312.4-320.1 | 111 | 111 |
| 25 | 269-282 | 320.3-328.0 | 181 | 177 |
| 26 | 283-296 | 328.1-335.8 | 136 | 136 |
| 27 | 297-309 | 335.8-342.5 | 111 | 111 |
| 28 | 350-352 | (29) 3.2-4.6 | 83 | 83 |
| 29 | 353-(08)001 | 4.7-11.4 | 429 | 426 |
| 30 | 002-012 | 11.5-16.7 | 260 | 258 |
| 31 | 016-026 | 18.2-23.2 | 123 | 123 |

| | | | | | | |
|---|---|---|---|---|---|---|
| 32 | 030-043 | 24.7-31.1 | 63 | | 77 | |
| 33 | 044-055 | 31.2-36.2 | 19 | | 21 | |
| 34 | 058-068 | 37.5-42.1 | 22 | | 22 | |
| 35 | 152-155 | 79.4-80.6 | 22 | | 22 | |
| 36 | 161-169 | 82.9-86.7 | 57 | | 57 | |
| **37** | **171-183** | **87.3-92.7** | **52** | | **52** | |
| **38** | **184-197** | **93.1-99.1** | **54** | | **54** | |
| **39** | **198-211** | **99.2-105.5** | **108** | | **108** | |
| **40** | **212-227** | **105.6-111.8** | **53** | | **53** | |
| **41** | **226-239** | **112.0-118.3** | **82** | | **82** | |
| **42** | **240-253** | **118.4-124.9** | **67** | | **67** | |
| **43** | **254-267** | **125.0-131.6** | **129** | | **129** | |
| **44** | **268-280** | **131.7-137.8** | **78** | | **78** | |
| **45** | **282-295** | **138.7-145.4** | **50** | | **50** | |
| **46** | **296-309** | **145.5-152.5** | **70** | | **70** | |
| **47** | **310-323** | **152.7-159.6** | **63** | | **63** | |
| **48** | **324-325** | **160.0-160.7** | **7** | | **7** | |
| **49** | **356-365** | **177.6-182.6** | **39** | | **38** | |
| 50 | (09)004-013 | 185.4-190.6 | 17 | | 17 | |
| 51 | 015-027 | 192.0-199.2 | 26 | | 26 | |
| 52 | 028-046 | 199.3-210.1 | 10 | | 10 | |
| No north observations during northern fall/winter (15 Feb-11 Jul 2009) | | | | | | |
| 64 | 199-202 | 305.3-307.2 | 3 | | 3 | |
| 65 | 211-217 | 312.5-315.9 | 3 | | 3 | |
| 66 | 224-237 | 319.9-327.5 | 45 | | 45 | |
| 67 | 238-238 | 327.6-327.7 | 4 | | 4 | |
| CRISM off due to MRO maintenance to 26 Aug-16 Dec, 2009 | | | | | | |
| 68 | 350-363 | (30)24.6-30.7 | 273 | | 274 | |
| 69 | 364-(10)012 | 30.7-37.0 | 525 | | 523 | |
| 69a | | | MSP | HSP | MSP | HSP |
| 70 | 013-026 | 37.3-43.4 | 52 | 536 | 52 | 535 |
| 70a | 027-040 | 43.4-49.5 | | 192 | | 192 |
| 70b | 041-054 | 49.7-55.7 | | 400 | | 404 |
| 70c | 055-068 | 55.8-61.9 | | 224 | | 225 |
| 71 | 069-082 | 61.9-68.0 | 67 | 216 | 67 | 157 |
| 72 | 083-096 | 68.3-74.1 | 15 | 200 | 15 | 200 |
| 73 | 097-110 | 74.1-80.2 | 12 | 314 | 14 | 313 |
| 74 | 111-124 | 80.2-86.4 | 209 | 235 | 188 | 224 |
| 75 | 125-138 | 86.4-92.4 | 190 | 62 | 189 | 62 |
| **76** | **139-149** | **92.6-97.2** | **89** | **8** | **89** | **8** |
| **77** | **154-166** | **99.6-105.0** | **30** | **154** | **30** | **154** |
| **78** | **167-180** | **105.2-111.4** | **115** | **0** | **115** | **0** |
| **79** | **181-194** | **111.5-117.9** | **95** | **0** | **95** | **0** |
| **80** | **195-207** | **118.0-124.0** | **52** | **0** | **51** | **0** |
| **81** | **231-236** | **135.5-137.9** | **2** | **108** | **2** | **108** |
| **82** | **237-250** | **138.0-144.9** | **538** | **106** | **539** | **109** |
| **83** | **251-256** | **145.0-147.9** | **141** | **10** | **141** | **18** |
| **84** | **285-292** | **163.1-166.8** | | | **35** | **35** |
| **85** | **293-306** | **167.0-174.5** | | | **67** | **67** |
| **86** | **307-312** | **174.6-177.7** | | | **67** | **67** |
| 87 | 337-348 | 191.7-198.1 | | | 8 | 8 |
| 88 | 349-353 | 199.1-201.1 | | | 4 | 4 |
| 89 | 363-001 | 207.3-209.5 | | | 2 | 2 |
| 90-95 | Nothing | | | | | |
| 96 | (11)177-179 | 317.4-319.2 | | | 28 | 28 |
| 97 | 180-183 | 319.2-321.4 | | | 41 | 41 |
| 98 | 226-232 | 344.6-347.7 | | | 14 | 14 |

Table 1 - Number of CRISM MSP Strips taken of Mars northern pole (defined as all strips partially or completely poleward of 55°N) per fortnight (starting at the TRAnsition cycle of the MRO primary science mission). Northern spring starts at $L_s = 0^\circ$ and ends at $L_s = 90^\circ$. MSP=multispectral polar observation, HSP=hyperspectral polar, MY = Mars Year, $L_s$ = solar longitude from Mars. DOY = Earth Day of Year **Bold text** indicates summer observations, which are not specifically discussed in this paper.

## $H_2O$ Ice Detection and Grain size Estimation Strategy

*$H_2O$ ice detection*. $H_2O$ ice dominates the north polar cap, and to map its presence in a CRISM pixel we use a $H_2O$ ice index first proposed by *Langevin et al.* [2007] and modified for CRISM mapping data by *Brown et al.* [2009]:

$$H_2Oindex = 1 - \frac{R(1.500)}{R(1.394)^{0.7}.R(1.750)^{0.3}} \qquad \text{(Eq. 1)}$$

where $R(\lambda)$ is the apparent reflectance at wavelength $\lambda$ in microns.

In our ice identification maps, we use determine that $H_2O$ ice is present if the $H_2O$ index is above a threshold value of 0.125, which we determined by iterative assessments of noise removal from our maps.

*$H_2O$ ice Grain size estimation*. In order to estimate the water ice grain size from the $H_2O$ ice index, we used a monodisperse, one dimensional approximate radiative transfer model to construct artificial $H_2O$ ice reflectance spectrum proposed by *Shkuratov et al.* [1999]. This model has a 'porosity' parameter lying between (0,1] that describes the volume fraction occupied by the target material and free space. A porosity value of 1 indicates no free space and a value of 0.5 indicates that half the volume is occupied by the target material.

The Shkuratov model produces as output a 1-d grain size parameter that may be imagined as an 'equivalent path length' for photons between scatterings, therefore the grain sizes we quote in this paper are actually fictionalized average photon scattering path lengths. These do not correspond to a spherical or cylindrical diameter or radius because the model is 1 dimensional and does not replicate spherical shapes. Therefore in this paper, the term 'grain size' is the same as the 'equivalent path length' of light within the model.

We used water ice optical constants of *Warren* [1984] and palagonite optical constants of *Roush et al.* [1991] (as a dust simulant) as input to the Shkuratov model. After carrying out the calculations at the resolution of the optical constants, we then convolved the model output spectrum to CRISM wavelengths before measuring the $H_2O$ index of the model spectra.

The dust grain size and volume percentage were varied in order to kept the visible albedo of the mixture around 0.5, which is typical of residual ice cap apparent visible albedos [*Kieffer*, 1990]. Here we consider 'visible albedo' to be equivalent to the apparent Lambert reflectance at ~0.8 $\mu$m. In order to match this albedo, we iteratively adjusted the volumes of our two-component model of $H_2O$ ice and dust. We found that a reasonable model resulted from $H_2O$ ice occupying 70% of the volume and dust grains (with a size of 35 microns) occupying 30% of the volume.

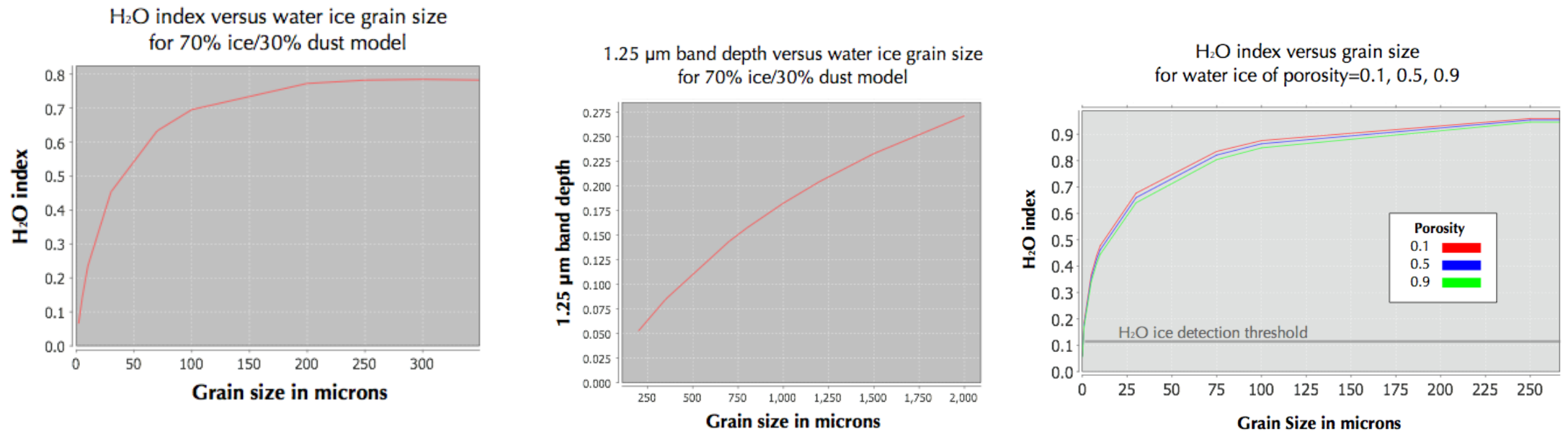

Figure 2 (a) Modeled relationship of particulate $H_2O$ ice index (Eqn 1) versus grain size for $H_2O$ ice grain size = 2 to 300 microns.
(b) 1.25 μm absorption band depth versus $H_2O$ ice grain size (200 to 2000 microns). Palagonite dust has been added to model to match visible albedo ~0.5. The model is 70% by volume water ice grains and 30% by volume 35 micron palagonite grains.
(c) Modeled relationship of particulate pure $H_2O$ ice index versus grain size for porosity of 0.1, 0.5 and 0.9. $H_2O$ ice index (reliant upon the 1.5 μm band depth, see Eqn 1) detection threshold of 0.125 is also shown.

*Model to convert $H_2O$ index to water ice grain size*. Figure 2a presents the curve for conversion of H2O index to water ice grain size we have used in this study. We have chosen to report the $H_2O$ index in our maps and histograms rather than grain size because grain size derivations require assumptions about the snowpack porosity, size distribution and an estimate of other components (e.g. dust). We consider it important to present estimated grain sizes, however there are caveats and assumptions in any model that infers grain sizes from reflectance spectra and we detail these in the paragraphs below.

To generate Figure 2a, we created the simplest spectral model we could imagine that can roughly fit the CRISM spectra observed. It is possible to imagine more complex scenarios, however our attitude to fitting infrared spectra follows the two principles of 1.) Occam's Razor and 2.) 'if you don't observe a unique set of absorption bands, make sure you clearly explain your deductive reasoning'. To this end, we used our two component $H_2O$ ice/dust model in an attempt to simulate a $H_2O$ ice snowpack with embedded (intimately mixed) dust grains. This two-component model is clearly an oversimplification of reality but we believe that it is necessary to adopt such a model in order to account for the relatively low visible albedo of the cap (pure $H_2O$ ice typically exhibits visible albedos of 0.99 [*Grenfell, et al.*, 1994] whereas, as stated earlier, the north polar cap typically exhibits visible albedos of 0.5).

Our approach has the advantage of calculating the reasonably straightforward 'minimum apparent grain size' [*Brown, et al.*, 2009]. Because we have chosen a two component (ice-dust) model, rather than a range of components or a size distribution, we are essentially calculating the smallest water ice grain size that will fit the observations. This means that we have made a positive detection of the minimum water ice grain size that is required to explain the observations – we can say with certainty that water ice of at least this grain size is present, otherwise the 1.5 μm water ice feature would not display this band depth. Use of the 'minimum apparent grain size' phrasing is not common in the radiative transfer literature, but we feel this terminology merits wider adoption given its simplicity and physical plausibility [*Brown, et al.*, 2009].

*Model for water ice grain sizes larger than 100 microns*. Beyond 100 microns grain size, the depth of the 1.5 μm band for our two-component model becomes saturated and it is therefore not suitable for grain size retrieval. For this reason, beyond a $H_2O$ index of ~0.5, if one is to obtain reliable grain size estimates, it is necessary to use an alternate water ice index. We have chosen to use the depth of the 1.25 μm band, which becomes prominent (greater than 0.02 relative band depth) for grain sizes larger than 100 microns. Figure 2b shows the curve for converting between 1.25 μm band depth and grain size.

To generate Figure 2b, we used the Shkuratov model described earlier (with a mixture of 70/30 percent by volume water ice/palagonite) to deduce the relationship between the 1.25 μm band depth and the water ice grain size. To estimate the amplitude of the 1.25 μm band depth, we used a band depth algorithm described in *Brown* [2006] and implemented in *Brown et al.* [2008b]. We discuss the implications of using the 1.25 μm band depth in this manner in the Observations and Discussion sections below.

*Errors induced by $H_2O$ snowpack density and porosity estimation*. In order to make an estimate of the errors induced by assuming a porosity of 0.5 in our spectral model, we have constructed models where porosity is 0.1 (fluffy snow), 0.5 and 1.0 (slab ice). These three models are shown in Figure 2c. Using our assumed porosity of 0.5, for $H_2O$ ice with a grain size of 10 microns, the range of induced error is ±1 microns, for grain sizes of 100 microns, the range is ±6 microns and for grain size of 250 microns, the range is ±12 microns. Based on these ranges, we can make an assessment that there is a relatively large induced error (up to ±6%) introduced if our porosity estimate is incorrect. The uncertainty introduced by noise is minimal, since CRISM SNR at 1.5 μm is ~400 [*Murchie, et al.*, 2007]. For a band depth of 0.08 this would equate to (1/400)/0.08=3.5%, much lower than our porosity error. Assuming the porosity and instrument errors are statistically independent, and using the relation *total error = sqrt(errorA$^2$+errorB$^2$)* then the instrument error gives an additional ~0.6% to the porosity error. Rounding up to one significant digit, we use ±10% for the relevant error estimate when quoting grain sizes later in the text.

## $CO_2$ Ice Detection Strategy

In order to detect and map $CO_2$ ice we used the $CO_2$ absorption band centered at 1.435 μm. To construct our ice identification maps, we determined that $CO_2$ ice was present if the 1.435 μm band depth exceeds a threshold value of 0.16 [*Brown, et al.*, 2009]. We have not attempted to derive maps or histograms grain size of seasonal $CO_2$ ice in this paper due to the challenges of separating $H_2O$ and $CO_2$ ice signatures (there is almost no 'pure' $CO_2$ ice in the maps we created here, in contrast with the south [*Brown, et al.*, 2009]. We report on changes of spectra of $CO_2$ ice at individual locations below in the Observations section.

### Seasonal Cap Retreat Line Mapping

As discussed in detail in *Brown et al.* [2009], we have devised an automated method to find the edge of the polar cap. In this paper, the 'Cap Recession Observations indicated $CO_2$/$H_2O$ has Ultimately Sublimated' (CROCUS/CROHUS) line corresponds to the most equatorward detection of $CO_2$/$H_2O$ ice. The term 'CROCUS' was first introduced by *Kieffer et al.* [2000] to describe the retreating edge of the $CO_2$ seasonal cap, the term 'CROHUS' is introduced here to describe the edge of the $H_2O$ ice seasonal cap. For comparison with previous observations utilizing only visible imagery to detect the edge of the cap, the edge of the visible cap corresponds to the CROHUS ($H_2O$) line and the CROCUS ($CO_2$) line lies poleward of this (since $H_2O$ ice is more stable than $CO_2$ ice to ambient temperature changes). This distinction is discussed in greater detail below.

We have constructed MARCI daily global mosaics in a similar manner to *Brown et al.* [2009]. For an in depth analysis of MARCI cloud and dust observations over a similar time period as this study, see *Cantor et al.* [2011].

## OBSERVATIONS

### Seasonal mosaics

The seasonal CRISM and MARCI mosaics are presented in Figures 3a-e. Alongside each image is the $L_s$ range of the CRISM observations, the number of CRISM mapping images used to construct the mosaic, and the $L_s$ corresponding to the MARCI image.

For Figure 3a-d, the left column displays a MARCI image, the central column displays the $CO_2$ ice-related 1.435 $\mu$m band depth and the right column displays the ice identification map for each time period.

Figure 3e differs from Figure 3a-d because instead of the 1.435 $\mu$m band depth maps in the central column, the $H_2O$ ice index (Eqn. 1) has been mapped. This is because no $CO_2$ ice signatures are apparent during these time periods. Figure 3e also includes an extra column (on the far right), showing the 1.25 $\mu$m band depth maps for each period for comparison with the $H_2O$ ice index.

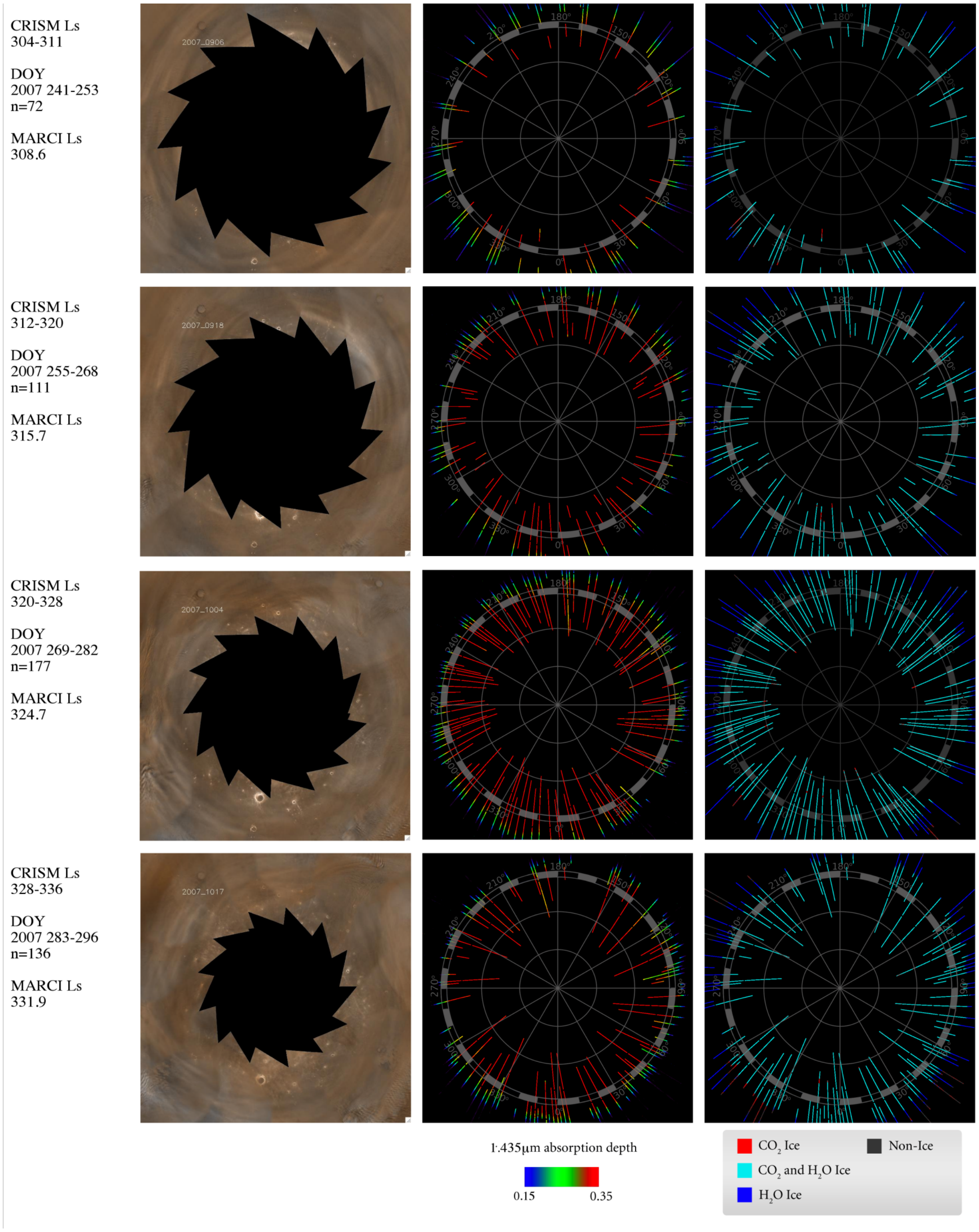

Figure 3 (a) North polar seasonal mosaics for period from MY 28 $L_s$=304°-336°. (left) Daily MARCI mosaic. (center) $CO_2$ 1.435 μm absorption depth (right) Ice identification maps. Polar stereographic map extends equatorward to at least 55°N. Latitude circles indicate 10° increments – 80°N, 70°N, 60°N are shown.

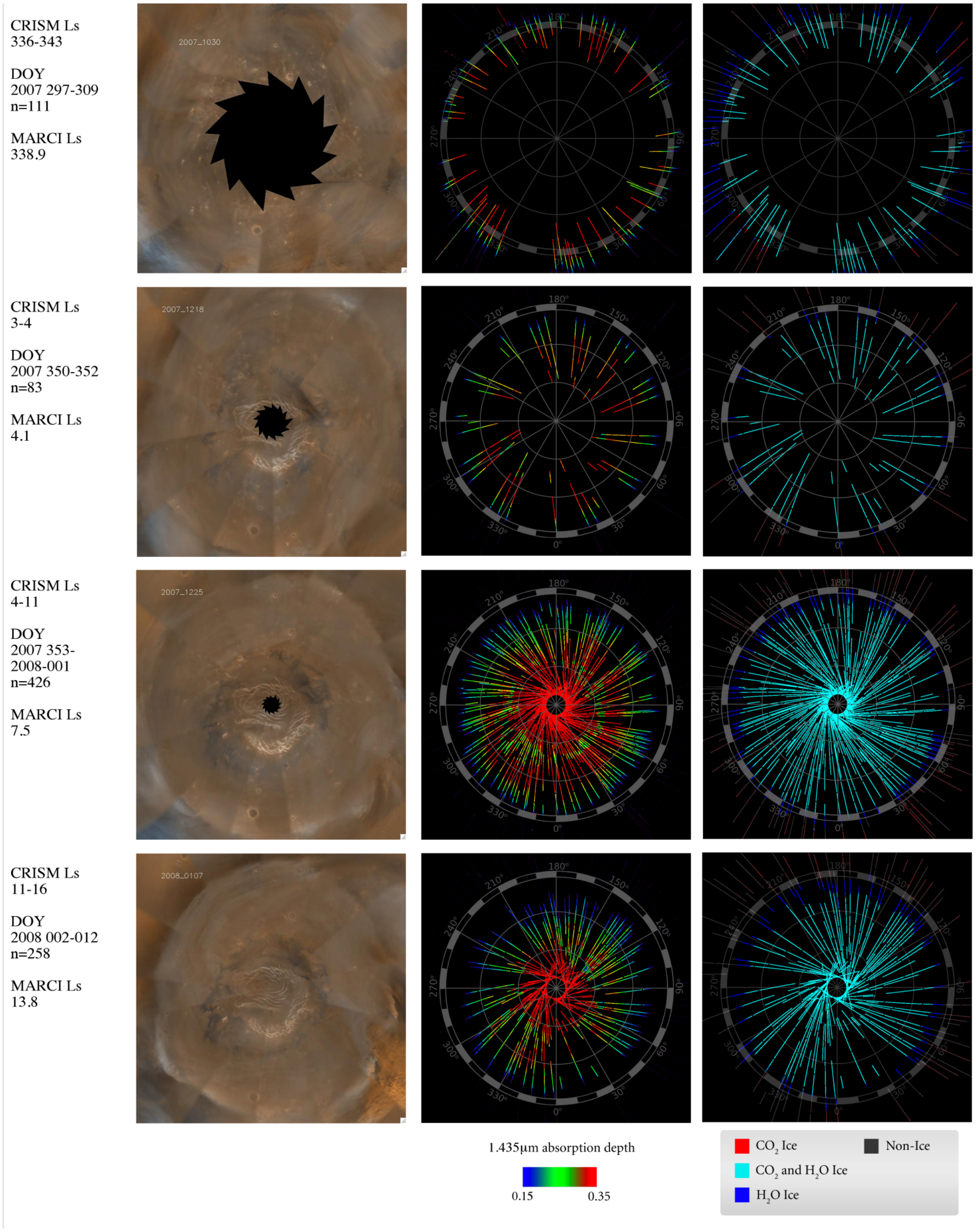

Figure 3.b. North polar seasonal mosaics for period from MY 28 $L_s$=336° - MY29 $L_s$=16°. (left) Daily MARCI mosaic. (center) $CO_2$ 1.435 µm absorption depth (right) Ice identification maps. Polar stereographic map extends equatorward to at least 55°N. Latitude circles indicate 10° increments – 80°N, 70°N, 60°N are shown.

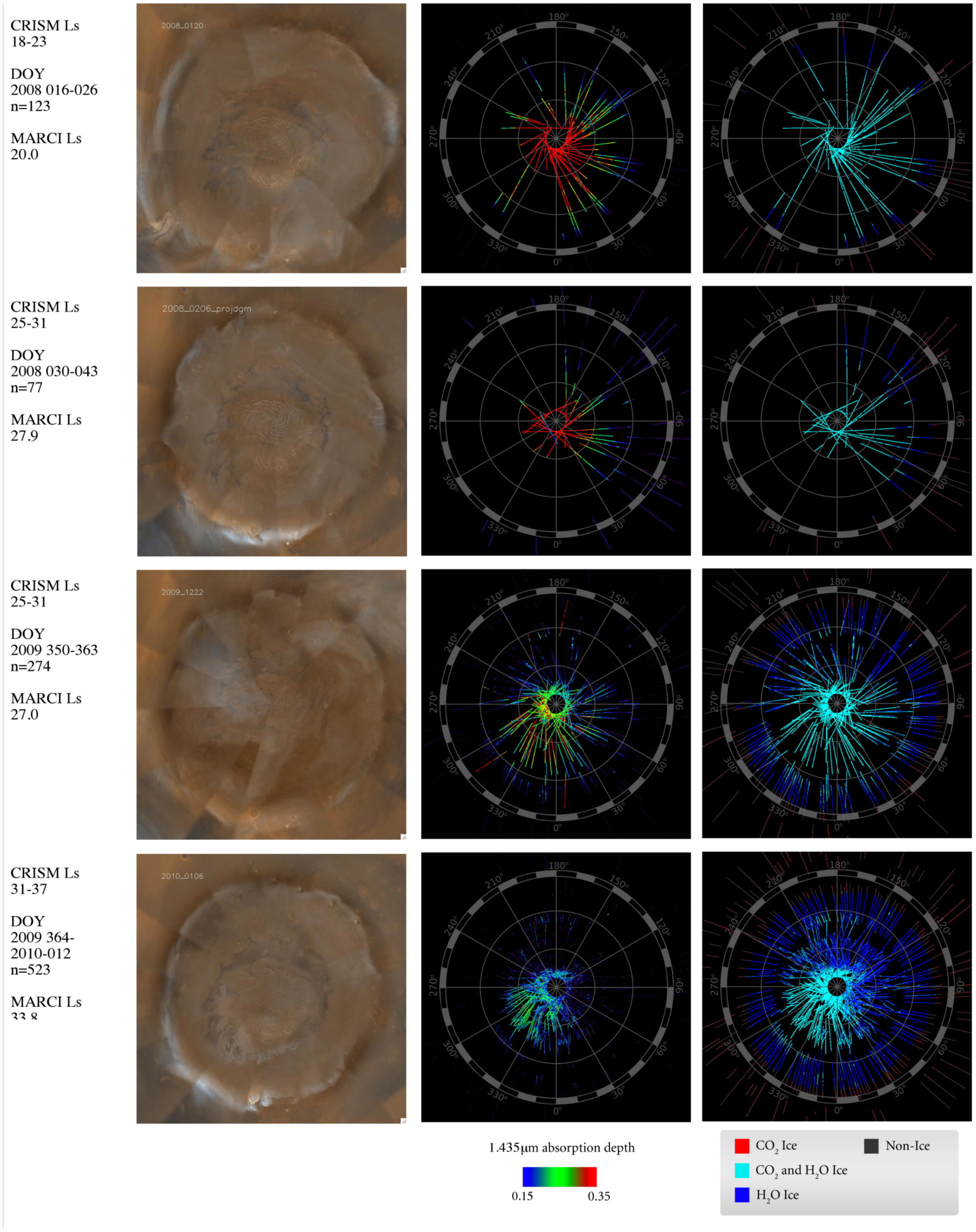

Figure 3.c. North polar seasonal mosaics for period from MY 28 and 30 $L_s$=18°-37°. (left) Daily MARCI mosaic. (center) $CO_2$ 1.435 μm absorption depth (right) Ice identification maps. Polar stereographic map extends equatorward to at least 55°N. Latitude circles indicate 10° increments – 80°N, 70°N, 60°N are shown.

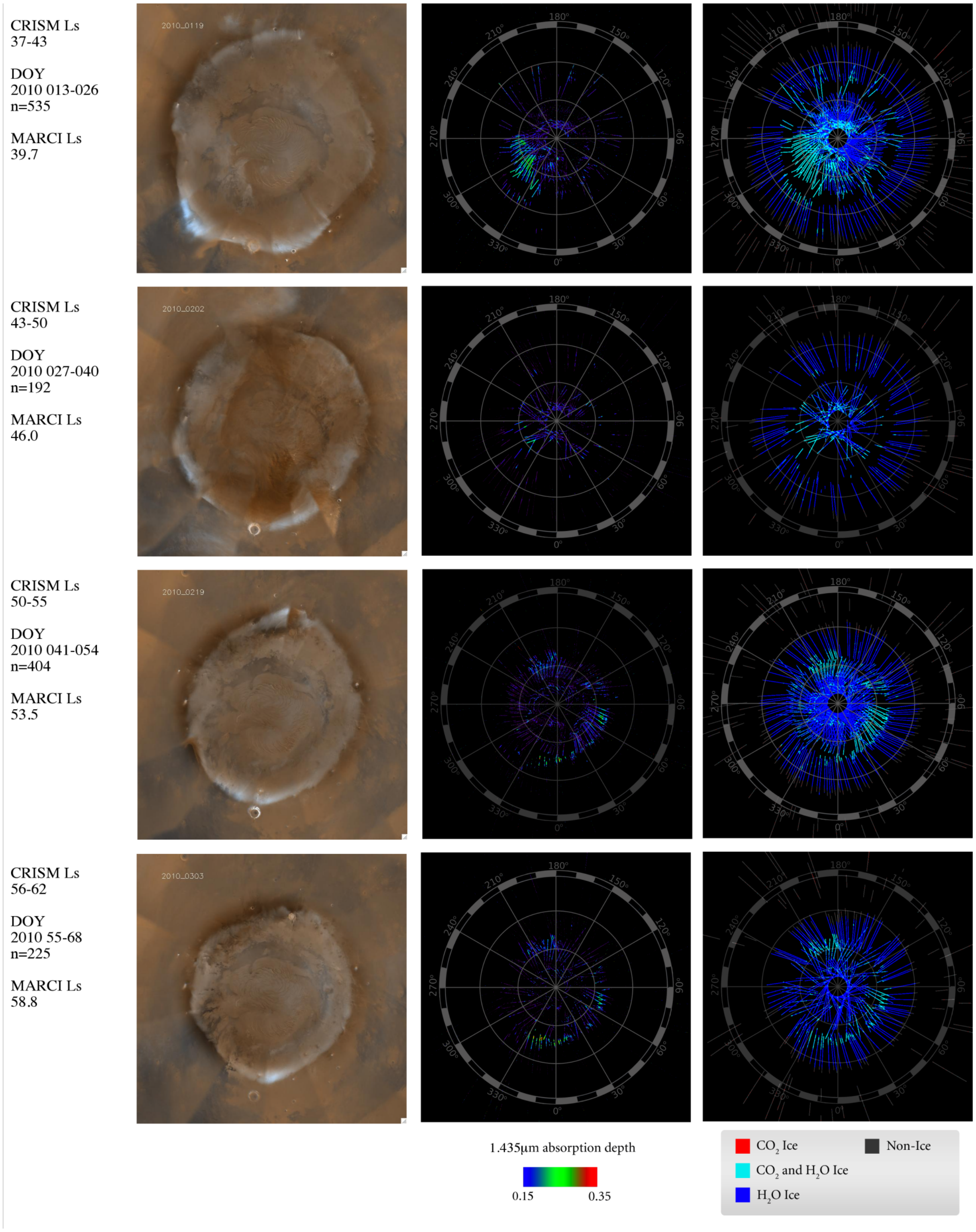

Figure 3.d. North polar seasonal mosaics for period from MY 30 $L_s$=37°-62°. (left) Daily MARCI mosaic. (center) $CO_2$ 1.435 μm absorption depth (right) Ice identification maps. Polar stereographic map extends equatorward to at least 55°N. Latitude circles indicate 10° increments – 80°N, 70°N, 60°N are shown.

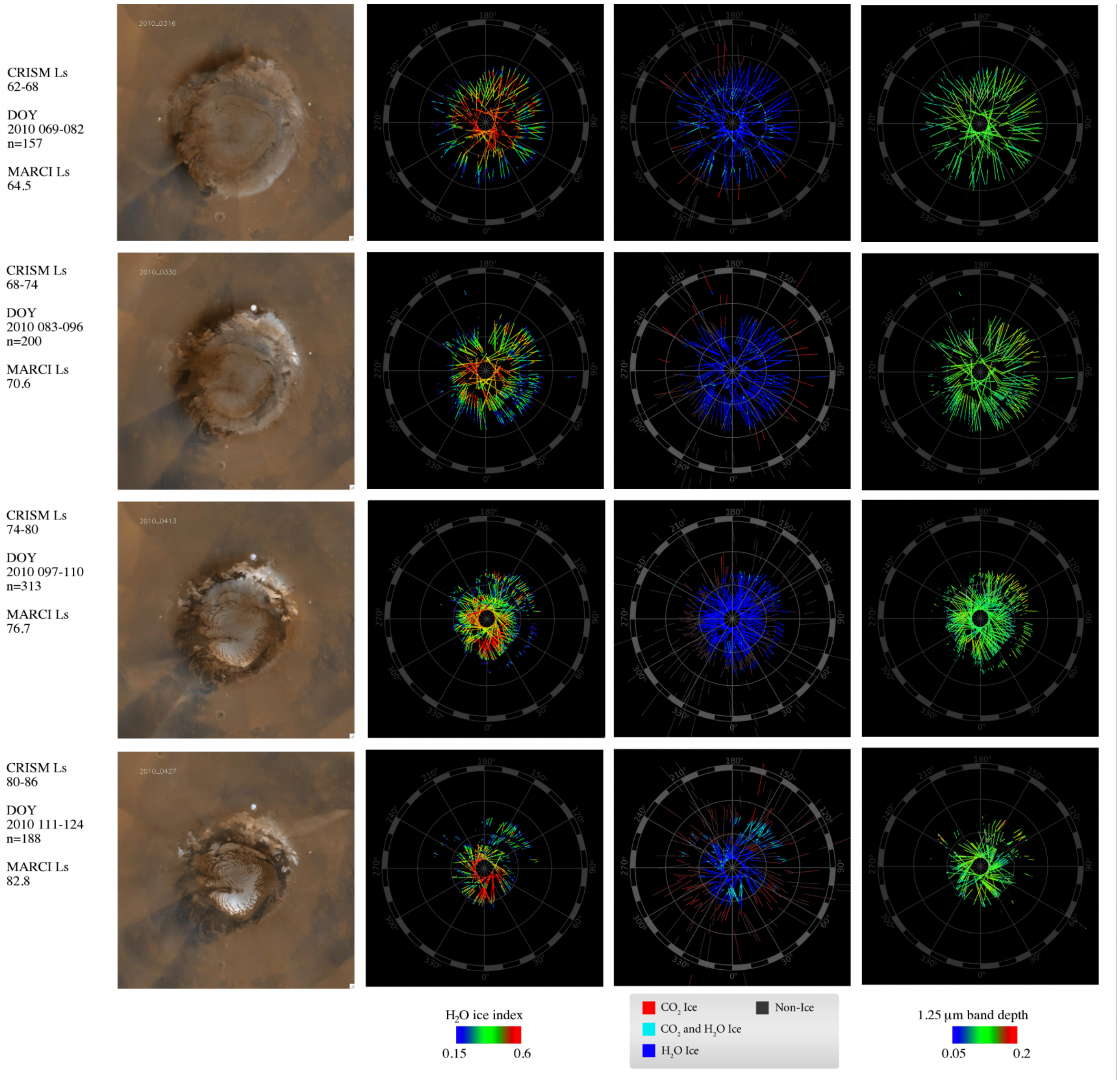

Figure 3.e. North polar seasonal mosaics for period from MY 30 $L_s$=62°-86°. (first column) Daily MARCI mosaic. (second column) $H_2O$ index (third column) Ice identification maps (fourth column) 1.25 μm absorption depth map. Polar stereographic map extends equatorward to at least 55°N. Latitude circles indicate 10° increments – 80°N, 70°N, 60°N are shown.

Figure 4 highlights the dynamic evolution of $CO_2$ and $H_2O$ ice by showing spectra from several different locations at different times in the cap recession. Figure 4a shows examples of coarse and fine grained $CO_2$ ice. The coarse grained example is taken from a mid-winter MSP observation of Utopia Planitia, and the fine grained example comes from the outer slope of Lomonsov Crater at the same period (they are ~0.1° $L_s$ apart). With a 1 μm albedo of around 0.85, the fine grained $CO_2$ ice spectrum is the brightest CRISM observation in the north pole that we have observed. Using grain size estimation methods in Figure 2b of *Brown et al.* [2009] the fine grain $CO_2$ ice has a grain size of 2 mm (2.28 μm

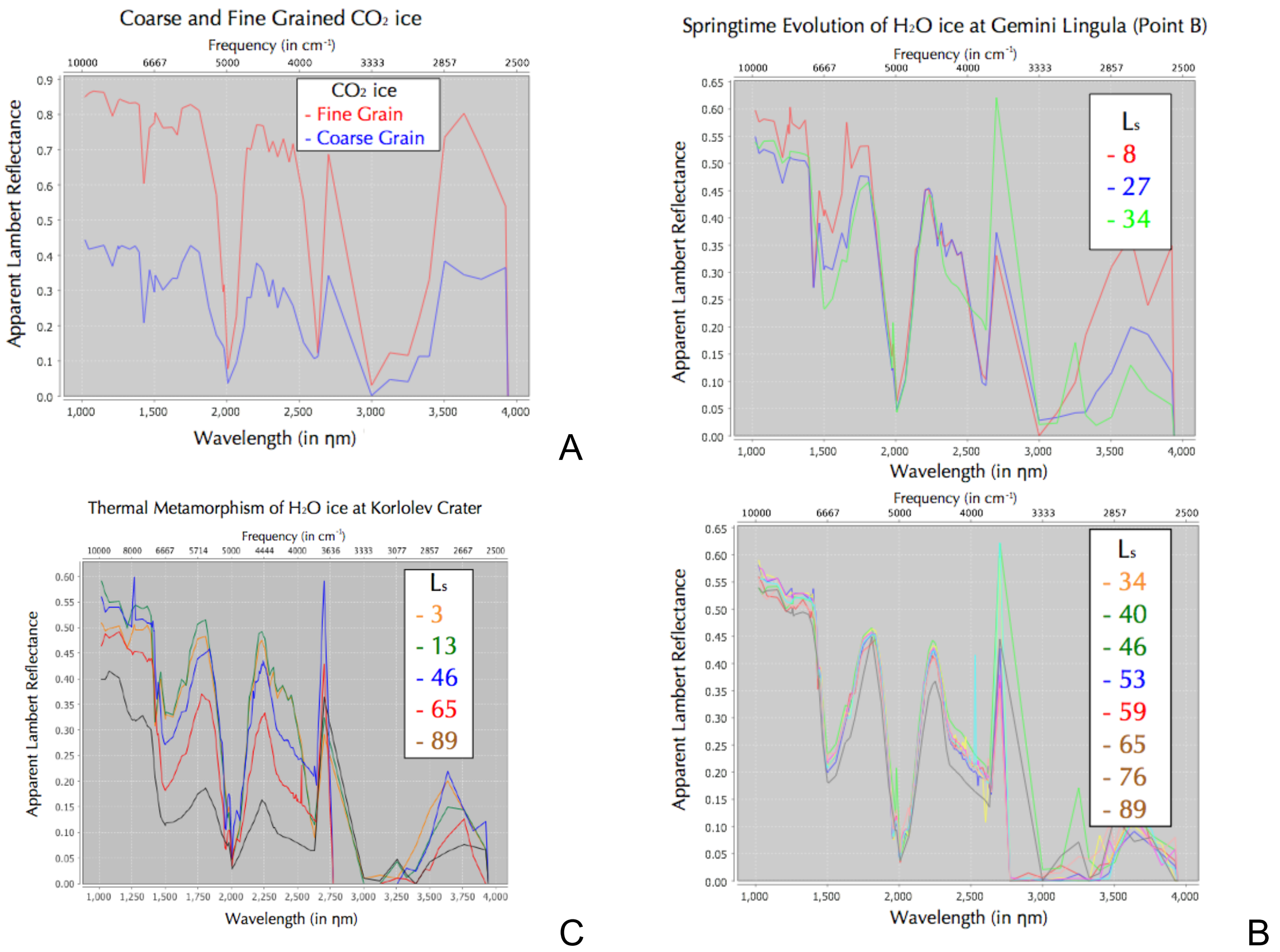

Figure 4. Representative spectra of spectral classes taken from the MSP dataset. See text for discussion. Reflectance here is "apparent Lambert reflectance" (incidence angle corrected) as discussed in the text and no atmospheric correction has been carried out.

a) 'Coarse-grained' $CO_2$ ice at 42.8°E, 66.2°N from CRISM MSP 787D_05 at $L_s$=309.6° on the plains of Utopia Planitia and fine-grained $CO_2$ ice at $L_s$=309.7° on the north slope of Lomonsov Crater at 348.9°E, 65.7°N from CRISM MSP 788B_01 $L_s$=309.7°. With a 1 μm albedo of around 0.85, this is the brightest CRISM spectrum detected in the polar regions to date.

b) Evolution of springtime ($L_s$=2°-89°) $H_2O$ ice grain signatures at 'Point B' [*Langevin, et al.*, 2005] on Gemini Lingula 45°E, 85°N (all spectra were collected from within 1 pixel of point 550, 550 on our 1000x1000 pixel mosaics).

c) Evolution of springtime ($L_s$=2°-89°) $H_2O$ ice grain signatures at 'Point C' [*Langevin, et al.*, 2005] on Korolev Crater, 164.2°E, 72.4°N (all spectra were collected from within 1 pixel of x=560,y=265 in 1000x1000 polar projection).

band depth of 0.08) and the coarse grained $CO_2$ ice spectrum from Utopia Planitia has a grain size of 10 mm (2.28 μm band depth of 0.22). We cannot rule out coarser or finer grained $CO_2$ being present in regions where CRISM did not collect observations (for example, the polar regions north of ~70°N were not observed during the winter period due to low light conditions).

Figure 4b and 4c present time evolution of polar cap spectra at two locations. Figure 4b presents observations from 45°E, 85°N on Gemini Lingula - 'Point B' of *Langevin et al.* [2005], who observed the water ice grain size growth from $L_s$=93°-127°. This region is also featured in Figure 7 of *Byrne et al.* [2008], where MOLA 1.064 μm albedos were obtained. This location is ideal for near-continuous observations due to the lack of topography and it is covered by ice year round. Figure 4b shows two springtime MSP L-channel spectra demonstrating that 'Point B' is covered by $CO_2$ ice at $L_s$=8° and $L_s$=27° (continuum-removed band depths of the 1.435 μm band are 0.45 and 0.39) then the $CO_2$ ice disappears at $L_s$=34° (1.435 μm band depth of 0.06). From $L_s$=34°-76° the water ice absorption bands are remarkably stable (with a $H_2O$ ice index of ~0.55 or around 50 microns from Figure 2a). *Langevin et al.* [2005] reported water ice grain growth beginning around $L_s$=93° – the spectra in Figure 4b show that ice grain growth actually commences around $L_s$=89°. The significance of what we have shown here is that the process is one of stagnation in ice growth for most of spring, followed by a large increase in ice growth starting around $L_s$=89°. This set of spectra therefore presents the strongest evidence for in situ thermal metamorphism of water ice in the Martian north pole to date.

Figure 4c presents the time evolution of CRISM spectra collected from 164.2°E, 72.4°N in Korolev crater, which is 'Point C' of *Langevin et al.* [2005]. Korolev has been shown to contain isolated regions of high thermal inertia which correspond to topography and have been interpreted as mounds of ice-rich regolith [*Armstrong, et al.*, 2005; *Armstrong, et al.*, 2007].

The spectra in Figure 4c show that $CO_2$ ice is present in Korolev during $L_s$=3° and 13° (continuum-removed band depths of the 1.435 μm band are 0.31 and 0.33), and then around $L_s$=46° the $CO_2$ ice is almost completely gone (1.435 μm band depth of 0.20). The $L_s$=46°, 65° and 89° observations show increases in the strength of the water ice bands. The $H_2O$ ice index for these spectra are 0.44, 0.55 and 0.56 respectively – the 1.25 μm band depths are 0.05, 0.04 and 0.14. This demonstrates that the 1.5 μm band has saturated between $L_s$=65° and 89°, as can be seen by the flattening of the 1.5 μm band in Figure 4c. This absorption band behavior is likely caused by rapid water ice grain growth (contrast this with the almost identical depth of the $H_2O$ ice absorption bands from $L_s$=34°-76° in Figure 4b). The faster increase in water ice absorption band growth of Korolev relative to 'Point B' is congruent with our interpretation of thermal metamorphism – water ice in Korolev is exposed to a warmer climate that will drive faster grain growth [*Gow*, 1969].

Comparison of the spectra at Point B and Point C demonstrates the inability of the $H_2O$ index (based on the 1.5 μm band depth) to discriminate effectively between water ice of grain sizes larger than ~100 microns. At $L_s$=89°, the Point B and C 1.5 μm band depths are very similar (0.58 and 0.56 respectively) however their 1.25 μm band depths are 0.04 and 0.14, demonstrating a true grain size of ~100 (Point B) and ~400 (Point C) microns according to Figure 2b.

## Seasonal Ice Cap Boundaries and Cap Area

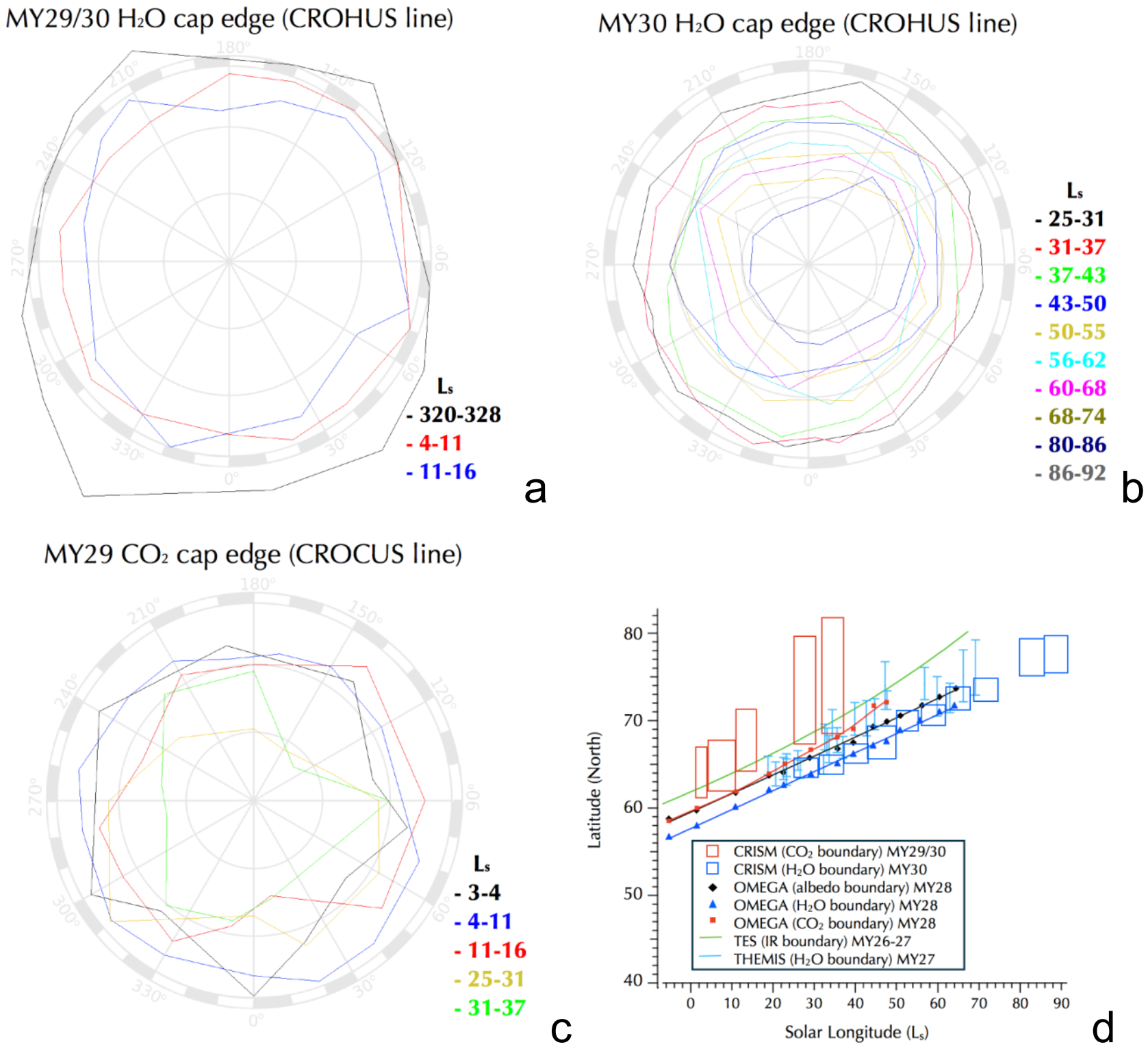

Figure 5 (a) $H_2O$ seasonal cap edge (CROHUS) lines for MY29 $L_s$=320°-16°. Polar stereographic map extends equatorward to at least 55°N. Latitude circles indicate 10° increments – 80°N, 70°N, 60°N are shown.
(b) $H_2O$ seasonal cap edge (CROHUS) lines for MY30 $L_s$=25°-92°. Polar stereographic map extends equatorward to at least 55°N. Latitude circles indicate 10° increments – 80°N, 70°N, 60°N are shown.
(c) $CO_2$ seasonal cap edge (CROCUS) lines for MY29 and MY30 $L_s$=3°-37°. Note asymmetry of cap for $L_s$=25°-37°. Polar stereographic map extends equatorward to at least 55°N. Latitude circles indicate 10° increments – 80°N, 70°N, 60°N are shown. See text for discussion.
(d) Comparison of CRISM MY30 $H_2O$ and $CO_2$ cap edge results to cap edge retreat rates of OMEGA [*Appere, et al.*, 2011], TES [*Titus*, 2005], THEMIS [*Wagstaff, et al.*, 2008]

Figure 5a and b plot the CROHUS line and Figure 5c plots the CROCUS line (defined using our ice identification thresholds mentioned earlier) as a function of time. Due to our interpolation method that is required to skip areas of missing data, the edge of the cap is unrealistically simplified, however our results are qualitatively similar to observations on previous years. Figure 5d shows the CRISM observations of the edge of the $CO_2$ ice cap and $H_2O$ ice cap from Ls=0°-90° for MY29 compared with observations for previous years [*Kieffer and Titus*, 2001; *Titus*, 2005; *Wagstaff, et al.*, 2008; *Appere, et al.*, 2011]. CRISM CROCUS/CROHUS boundaries are given as ranges covered by a box in order to illustrate the range of minimum and maximum latitude of the cap edge. It can be seen that the $CO_2$ cap edge boxes are very wide during the asymmetric retraction period. This feature was not reported in previous studies. The CRISM cap edges have been constructed automatically and we believe this accounts for the slightly faster retreat rates when compared to other observations, otherwise the comparisons show the cap retreat is relatively similar to other years.

In order to calculate the area of the cap, we used the polygons in Figure 5, and applied an algorithm for computing the area of an irregular polygon available in the open source Computational Geometry Algorithms Library (CGAL). We show the estimate of the cap area in Figure 6. If can be seen that the cap covers just less than 8% of the Martian surface at the mid-winter period and decreases to its summertime area around $L_s$=90°, when it covers ~1.4% of the Martian surface.

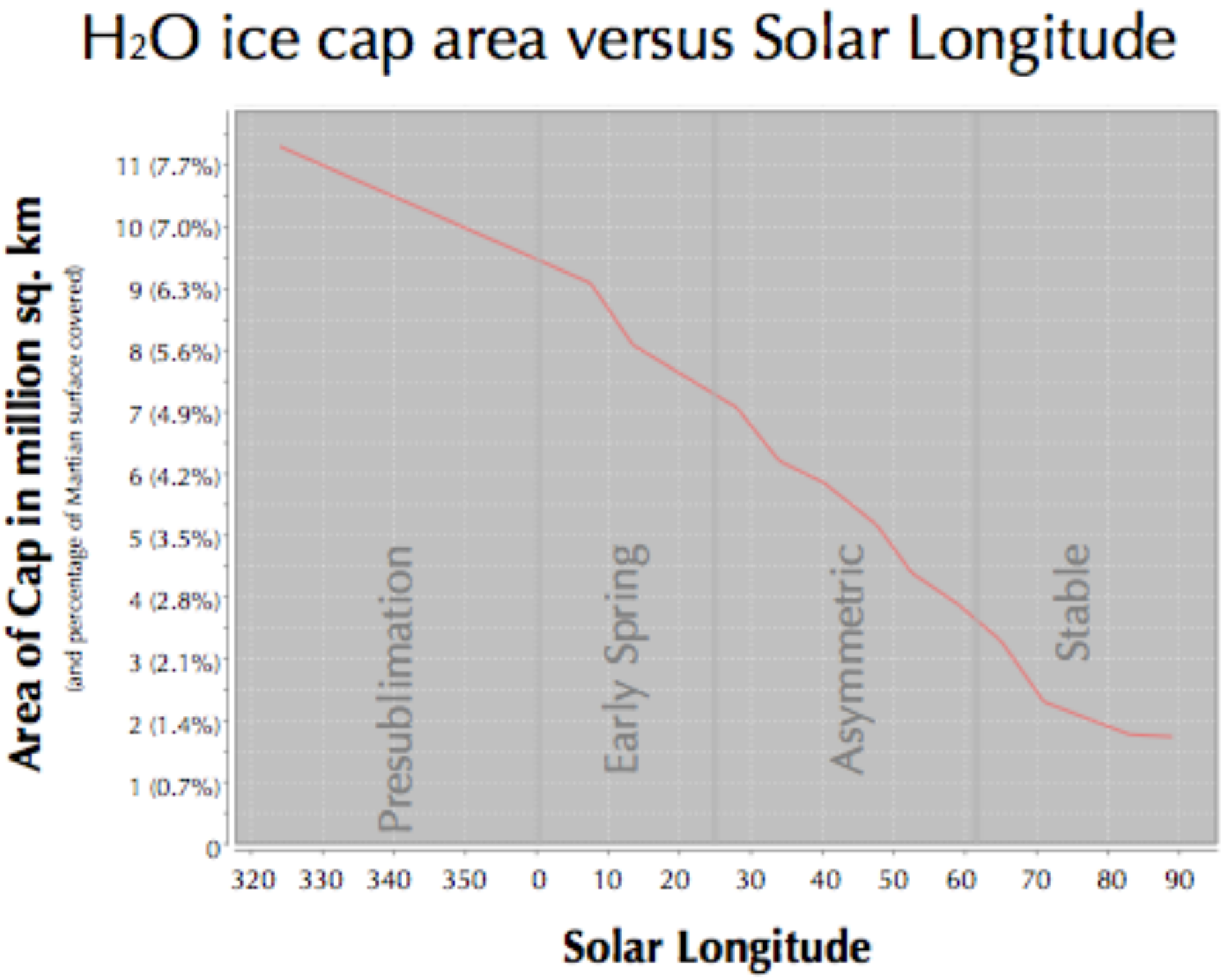

Figure 6. $H_2O$ seasonal cap area for $L_s$=320°-90°. Sublimation phases indicated are discussed in the text.

## Histograms of seasonal $H_2O$ ice grain size

The histograms of seasonal water ice abundances for each set of CRISM observations are presented in Figures 7a-e. Figures 7a-d use the $H_2O$ index and Figure 7e uses the 1.25 μm absorption band depth. The histograms have 100 bins and are normalized by the number of observations containing $H_2O$ ice. In Figure 7a-d, all pixels were included in the histograms. In Figure 7e, all pixels having a $H_2O$ index of greater than 0.125 were included in the histograms, even if those pixels also contained $CO_2$ ice. The histograms show that $H_2O$ ice indexes increase throughout the springtime, allowing us to divide the cap retreat into several time periods as discussed below.

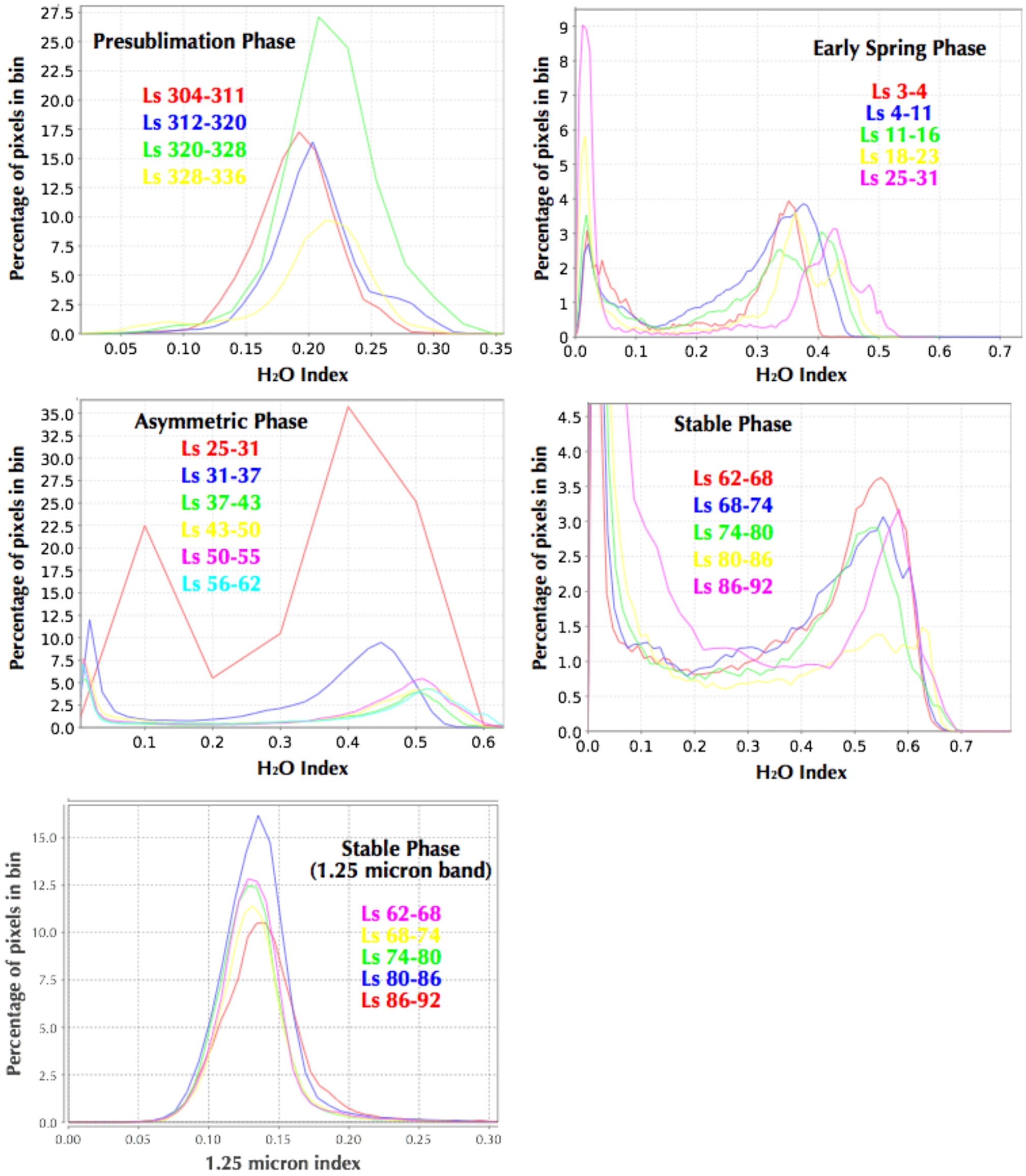

Figure 7a-e Histograms showing $H_2O$ ice grain size distributions. Note (d) and (e) cover same time period – (d) shows $H_2O$ index, (e) shows 1.25 μm band depth. See text for discussion.

# DISCUSSION

## Interpretation of Seasonal Mosaics

Examination of the CRISM seasonal maps leads to a natural division of the springtime recession into four different phases – presublimation, early spring, asymmetric retraction, and stable phases. The first phase, from $L_s$=304°-0° we term *presublimation* phase. From $L_s$=0°-25°, is the *early spring* phase, from $L_s$=25°-62° is the cap *asymmetric retraction* phase and from $L_s$=62°-92° is the *stable* cap phase.

*Apparent $CO_2$ ice cap*. In the following discussion we refer to the 'apparent $CO_2$ ice cap' in order to indicate those areas where CRISM detects $CO_2$ ice (according to the definitions above). As discussed further below, that does not match the TES $CO_2$ ice temperature map [*Kieffer and Titus*, 2001], which extends equatorward of the CRISM apparent $CO_2$ ice cap, indicating that $H_2O$ ice (most likely transported and cold-trapped by the Houben process) is obscuring a layer of underlying seasonal $CO_2$ ice that is responsible for the observed TES temperatures [*Wagstaff, et al.*, 2008; *Appere, et al.*, 2011].

*Presublimation phase ($L_s$=304°-0°).* Figures 3a-b cover this phase. This phase embodies the winter state of the Martian northern pole. The seasonal cap clearly extends equatorward of 50°N and is roughly symmetric around the pole at $L_s$=304°. Around $L_s$=330°, the retreat of the apparent $CO_2$ cap seems to slow along the 0°E line. By $L_s$=340°, the retreat is roughly symmetric again, and the apparent $CO_2$ ice cap edge is close to 58°N. The MARCI images show the seasonal cap has relatively low visible albedo during this time.

*Edge of the Apparent $CO_2$ ice Cap signatures*. At the edge of the $CO_2$ (mixed with $H_2O$) seasonal cap, we observe a decrease in the strength of the 1.435 μm $CO_2$ absorption band, as shown in the middle column of Figure 3a-b by blue and green pixels at the edge of the cap in the middle column. This decrease in 1.435 μm $CO_2$ band strength at the edge of the cap could be due to 1.) a decrease in area covered by the $CO_2$ cap (at fractions of the ~187.5 m scale) or 2.) obscuration of the $CO_2$ ice by $H_2O$ ice, or 3.) a decrease in the grain size as the edge of the cap is sublimating. These three effects are not separable without high resolution images of the cap edge, and in fact they may all act in concert to produce the effect observed. The edge of the cap thus affected by this decrease in the $CO_2$ ice absorption band forms an annulus around 10-100 km wide around the retreating cap.

*Pure $CO_2$ surface ice deposits in Lomonosov Crater between $L_s$=304°-336°*. In contrast to the south pole, where 'pure' $CO_2$ ice deposits (i.e. no $H_2O$ ice absorption bands present in a spectrum) are common, Figure 3a shows that it is extremely rare to find 'pure' $CO_2$ ice in the north pole, even in winter time observations. The exception is in Lomonosov crater at (at 65°N, 350°E) where pure $CO_2$ ice is present from $L_s$=304°-336° (see also Figure 4a). This is the only region that CRISM detected pure $CO_2$ ice over this mid-to-late winter time, although observations poleward of 70°N were not carried out in mid-winter due to low light conditions. The fact that the 'pure' $CO_2$ observations were sharply limited to the crater and present for such an extended period suggests that the crater may enjoy its own microclimate during this period which inhibits $H_2O$ ice deposition (or encourages $CO_2$ ice deposition).

*Decreasing width of $H_2O$ ice annulus*. At $L_s$=304°-311°, the earliest observations we have available, the $H_2O$ ice annulus can be seen to extend to (Figure 3a, third column) 47°E, 47°N. We checked a CRISM MSP observation taken at $L_s$=309°, (MSP 787D_01), that is not in Figure 3a, but extends further south. This observation shows that the true extent of the water ice annulus extends to 48°E, 44.3°N at $L_s$=309°. The $CO_2$ ice signature extends at this time to 46°E, 55°N, thus the water ice annulus spans over 10.7° of latitude at this time.

*Early Spring phase ($L_s$=0°-25°)*. Figures 3b-3c cover this phase. The middle columns of these images clearly show the rapid retreat of the $CO_2$ ice signatures from ~62°N to ~70°N. The exterior edge of the seasonal cap, as estimated by CRISM $H_2O$ ice signatures, retreat far more slowly, from ~59°N to 62°N, during the same period.

The seasonal cap $CO_2$ ice distribution during the Early Spring phase has an inhomogenous distribution, particularly apparent during the well-covered two week period of MY29 $L_s$=4°-16°. Deeper $CO_2$ ice-related 1.435 μm signatures are present on the north pole residual cap (NPRC) and surrounding bright outliers, but not the dark dunes that surround the NPRC, since these will be warmer during this period. A particularly thin covering of $CO_2$ ice can be seen on the dunes at 81°N, 122°E from $L_s$=11°-23°.

The width of the $H_2O$ ice annulus (Figure 3b, third column) decreases markedly in this period - by $L_s$=4°-11° it has decreased to around 2° of latitude in width.

*Asymmetric Retraction phase ($L_s$=25°-62°).* Figures 3c-d cover this phase. During this phase, $CO_2$ seasonal cap begins an 'apparent' asymmetric retreat. We use the term 'apparent' because temperatures measured by TES indicate no asymmetry around the cap [*Kieffer and Titus*, 2001], suggesting that the cause of the apparent asymmetric retreat is an asymmetric deposition of $H_2O$ ice onto the $CO_2$ cap, thus obscuring its spectral signature.

Figure 8 shows a simplified diagram of the suggested explanation for the asymmetric water ice distribution [*Brown, et al.*, 2011; *Mellem, et al.*, 2012]. This aeolian process releases more water in the longitude range 90-210°E, and is potentially due to exposure of rough residual ice outliers (the outliers are termed 'Mrs. Chippy's Ring' by *Calvin and Titus* [2008]).

Figure 8 shows the four stages that would lead to exposure of the water ice of the rough water ice outliers. In the first stage, a mid-winter scene shows $CO_2$ ice entirely covering the deposits. In the second stage, the $CO_2$ ice has begun subliming but the rough water ice deposits are not yet exposed. In the third and crucial stage, the water ice outliers are exposed, water vapor is formed and rapidly transforms to water ice grains that are then blown on top of the water ice, covering up the signature of $CO_2$ ice around them, and creating the asymmetric distribution of water ice that is observed. In the final stage, showing a summer scene, the water ice deposits are likely to be subliming and losing mass continuously. It should be emphasized that this process is subtly (but manifestly) different from the Houben process, which hypothesized sumblimation of the water ice annulus and cold trapping of gas molecules as the transport mechanism. *Houben et al.* [1997] describes the Houben effect in this manner (p. 9077) “the cap edge is also the location of intense baroclinic wave storm activity [Leovy, 1973; Barnes, 1981]. The warm phases of these eddies (which pick up most of the water vapor) are associated with poleward winds. They therefore transport the bulk of the water to higher latitudes. Our computations show that this newly released moisture subsequently re-precipitates from the cold phase of the baroclinic waves and accumulates at higher latitudes.”

However, for the model in Figure 8, the water is still in thermal contact with $CO_2$ ice and will not sublime, therefore the deposition in this model is necessarily driven by the wind as an aeolian process. The water ice that asymmetrically covers the $CO_2$ ice in our model does not travel in the vapor phase. This makes this process different from the Houben-inspired models described by *Bass and Paige* [2000], *Schmitt et al.* [2006] and *Appere et al.* [2011].

The $H_2O$ ice seasonal cap continues to shrink symmetrically around the pole during this time, retreating from ~58°N to just inside the 70°N circle by $L_s$=62°.

*Late Increase in $CO_2$ Signature (LICS) Events*. *Langevin et al.* [2008] and *Appere et al.* [2011] reported OMEGA observations of ‘Late Increases in $CO_2$ ice Signatures’ (LICS) around $L_s$=60°-80° and CRISM data shown here corroborates that finding. For example, Figure 3d shows $CO_2$ ice signatures in the middle of $H_2O$ ice signatures around 75°N, 340°E and 77°N, 80°E at $L_s$=50°-68°. *Appere et al.* attributed these $CO_2$ ice signatures to exposures of underlying $CO_2$ ice as $H_2O$ ice that had been obscuring the $CO_2$ was removed, perhaps by katabatic winds [*Spiga, et al.*, 2011]. The conditions under which these processes occur clearly require further laboratory and numerical investigation.

*Stable phase ($L_s$=62°-92°).* Figure 3e covers this phase. Since the ‘apparent’ $CO_2$ ice signatures have disappeared at this time, we show the $H_2O$ ice index images in the second column of this figure. These show the distribution of $H_2O$ ice is inhomogenous, and just as for the $CO_2$ ice cap, the weakest $H_2O$ index regions are over the relatively warm, dark dune regions surrounding the NPRC. This is outstandingly illustrated by the CRISM $L_s$=74°-80° images which show stable $H_2O$ ice south of the dunes in the 0-90°E radial, and no $H_2O$ ice on the dunes poleward of this region. MARCI images show Shackleton’s Grooves (80°E, 90°E) have been covered by dust at $L_s$=80°-86° [*Cantor, et al.*, 2011], and CRISM $H_2O$ ice signatures are accordingly weakened in that image.

On the fourth column of Figure 3e, the 1.25 μm band absorption depth is plotted. The interesting observation that becomes apparent from these 1.25 μm band maps is that the largest band depths are on the periphery of the cap and they grow throughout the Stable phase. This was not apparent from the $H_2O$ index alone, for reasons discussed earlier. The largest grain sizes correspond to Korolev and the water ice outliers ('Mrs. Chippy's Ring'), which strengthens the argument that the spectra in Figure 4c are revealing $H_2O$ ice metamorphism which is more rapid in the equatorward (warmer) ice deposits (e.g. those at 150°E, 75°N).

*Comparison to TES observations of MY 24-25 recession*. *Kieffer and Titus* [2001] showed that the thermal infrared coverage by TES of the springtime recession was symmetric and that stable $CO_2$ ice temperatures completely disappeared from the cap at $L_s$=78°. This slower, symmetric retreat contrasts with the asymmetric 'apparent' $CO_2$ ice cap retreat observed by CRISM. This supports the suggestion that cold-trapped $H_2O$ ice is obscuring the edge of the $CO_2$ ice cap [*Houben, et al.*, 1997; *Bass and Paige*, 2000; *Schmitt, et al.*, 2006; *Wagstaff, et al.*, 2008; *Appere, et al.*, 2011]. We have shown here that this process is asymmetric after $L_s$=25°, although the physical process responsible of this asymmetry is not well constrained and we have suggested only a simple model of this process here (Figure 8). The nature of the daily wind regimes and $H_2O$ condensation processes clearly requires further laboratory and numerical study.

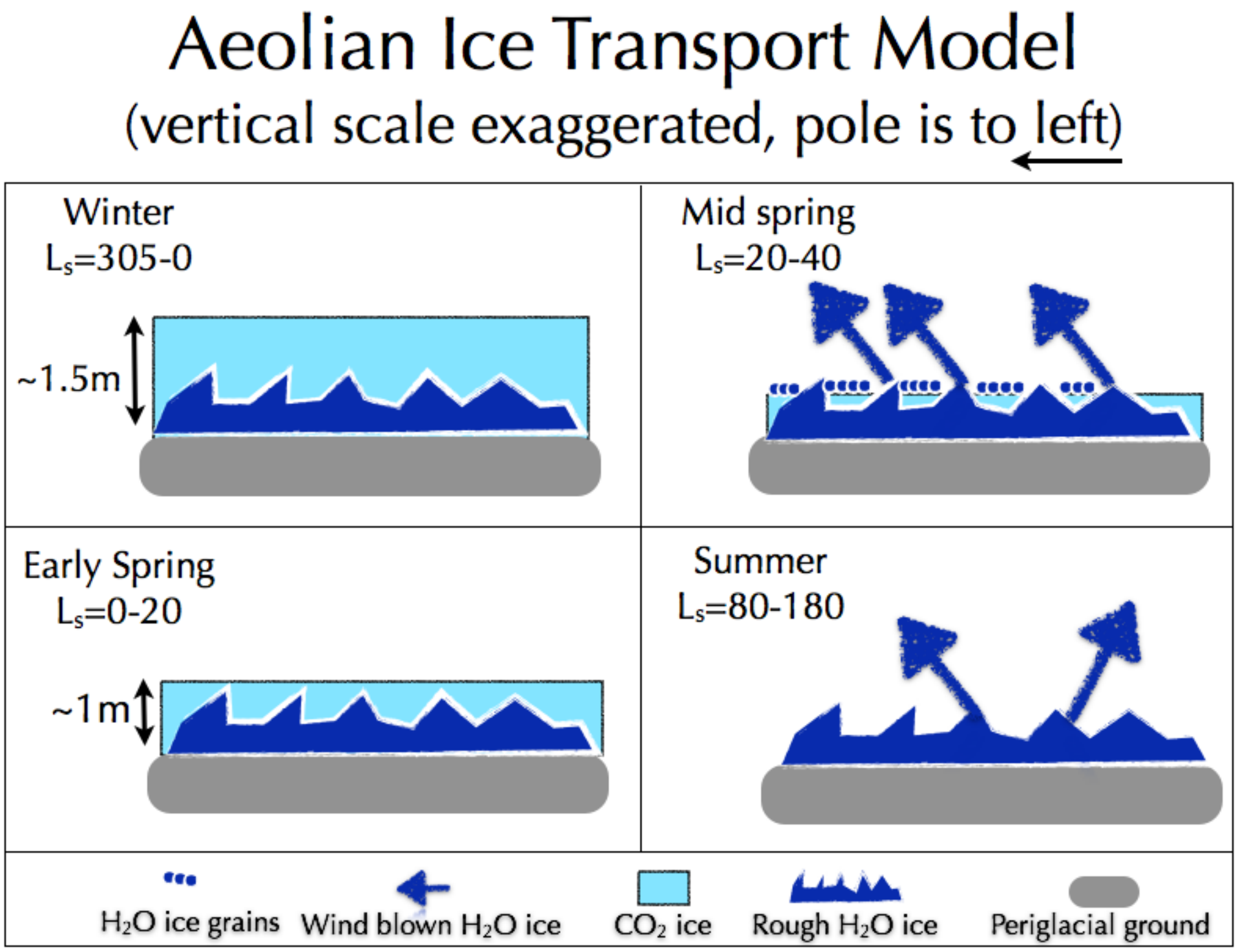

Figure 8. A simplified diagram of the proposed aeolian $H_2O$ ice transport process responsible for the asymmetric water ice distribution in the mid-spring ($L_s$=20°-40°). See text for discussion.

*Comparison to OMEGA observations of MY27-28 recession*. *Appere et al.* [2011] observed the retreat of the cap using OMEGA data, and as has been reported earlier, revealed the presence of LICS events in late spring. *Appere et al.* noted some asymmetry in the early spring period in the 210-270°E (compared to 330-030°E) radial (their Figure 15) and suggested albedo, $H_2O$ ice and $CO_2$ ice all retreated faster along the 210-270°E radial. They suggested this may be due to topographic differences or weather patterns [*Hollingsworth, et al.*, 1996]. Our results suggest the asymmetry is only in $CO_2$ ice and possibly albedo (see Figure 3c, last row) and the slow $CO_2$ ice retreat is from 270-30°E leading us to conclude that exposed rough ice in the ice cap outliers may be transporting material to other regions of the cap (Figure 8).

*Appere et al.* [2011] also observed that the $H_2O$ ice annulus extent was relatively large at in the mid-to late winter (they observed it was 6° wide at $L_s$=330°) and decreased in width (they observed it was 2° wide at $L_s$=350°). This roughly corresponds to CRISM observations of the span of the $H_2O$ ice annulus prior to $L_s$=25° reported here. Some differences between the OMEGA study and the CRISM data reported here may be due to interannual variability.

**Interpretation of $H_2O$ Grain Size Results**

Histograms showing the water ice index are shown in Figure 7a-d. A histogram of the 1.25 μm band depth is shown in Figure 7e. The histograms are divided into four different phases as discussed below.

*Pre-sublimation phase $H_2O$ ice grain sizes*. From $L_s$=304°-336°, mid-winter of MY 28, the range of $H_2O$ ice index is relatively restricted (Figure 7a). During this period, the largest $H_2O$ ice index of 0.35 was achieved at $L_s$=320°-328°, however it is closely matched by the $L_s$=328°-336° observations. The smallest peak modes of around 0.2 are for the earliest periods, $L_s$=304°-320°. Assuming they are due to $H_2O$ and dust alone and our models (Figure 2a) are correct, this would correspond to minimum apparent grain sizes slightly less than 10±1 microns.

*Early Spring phase $H_2O$ ice grain sizes*. As spring begins to warm up the north polar region (from $L_s$=3°-25°, MY28) the $H_2O$ index increases markedly, quickly adopting values roughly from 0.2-0.5, with modes around 0.35-0.42. Interpreting these values using Figure 2a indicates they correspond to a grain size range of 10-40 microns, with modes of 30-35 microns.

*Asymmetric Retraction phase $H_2O$ ice grain sizes*. During $L_s$=25°-62°, MY30, $H_2O$ index profiles for polar ice continue to increase. The maximum values exceed 0.6 and minimums move up to around 0.3 (corresponding to a grain size maximum of 60±6 microns and minimum of 20±2 microns). The modes for the last four periods ($L_s$=37°-62°) are tightly bound around 0.5 (corresponding to grain sizes of 40±4 microns).

*Stable phase $H_2O$ ice grain sizes*. During the late spring stable period, the relative band depths are also the most stable and constricted. The $H_2O$ indexes have all increased above the values in the asymmetric period. Practically all of the $H_2O$ ice index values are between 0.7 (grain size of 100±10 microns) and 0.4 (grain size of 30 microns). The $H_2O$ ice index mode is around 0.55 (corresponding to a grain size of 50±5 microns).

As discussed earlier, because the $H_2O$ index range of the Stable phase (late spring) exceeds 0.6, we consider it unreliable as an indicator of water ice grain size. In Figure 7e we have plotted the 1.25 μm band depth for five periods of the Stable phase. The values range from a minimum of 0.075 (corresponding to around 300±30 microns from Figure 2b) to just over 0.2 (corresponding to 1200±120 microns), with a mean of around 0.13 (corresponding to a grain size of 600±60 microns). These grain sizes are well in excess of those estimated by the $H_2O$ index plot (Figure 2a) but we believe they are closer to reality due to the severe damping of the 1.5 μm band when the water ice grain sizes become this large.

*Interpretations of evolution of $H_2O$ ice grain sizes*. During the $L_s$=304°-92° mid-winter to springtime period, $H_2O$ ice grain sizes are continuously growing. This is in contrast to the $CO_2$ ice grain sizes in the south polar residual cap reported by *Brown et al.* [2009]. The increase in grain sizes might be explained in two ways: 1.) by sublimation of fine grained ice and exposure of large grained $H_2O$ ice or 2.) by metamorphism/sintering of $H_2O$ ice grains during this period. Both options could explain our observations, and both these processes require further laboratory and numerical study.

*Comparison with previous grain size investigations*. *Kieffer* [1990] used a simple $H_2O$ ice metamorphism model to predict that if the north polar cap is undergoing net annual sublimation then late summer grain sizes would grow to 100 microns or more. *Clow* [1987] predicted $H_2O$ Martian ice grains of 50 microns in size may grow to 1000 microns in size in the mid-latitudes by thermal metamorphism or sintering in one summer season.

*Langevin et al.* [2005] reported observations of the growth of $H_2O$ grains in the north polar cap for early summer ($L_s$=93°-127°). Their study plotted spectra for three locations and used radiative transfer models to suggest sub-100 micron grains at $L_s$=93° were being replaced by 700-800 micron sized grains at $L_s$=127° at the Point B locality. They suggested 700 micron-1 mm water ice grains were present at Point C in Korolev at $L_s$=93.6°. Our estimates are largely similar, however we report slightly larger grains (with a maximum of 1.2 mm) because we are using the 1.25 μm band depth method devised for this study. *Langevin et al.* suggested thermal metamorphism could not be responsible for this growth (~100 micron to ~1 mm in 1 Earth month) and suggested that fine grained ice was being sublimated away during this period to reveal older, coarser grained ice beneath. CRISM northern summertime results will be the subject of future study.

*Appere et al.* [2011] reported modeling 200 micron $H_2O$ ice grain deposits on top of $CO_2$ ice deposits in order to obscure underlying $CO_2$ ice signatures when $CO_2$ grains were 7 cm, however as they acknowledged, determining the grain size of $CO_2$ ice in the presence of large amounts of water ice is a difficult challenge because the broad $H_2O$ 1.5 $\mu$m and narrow $CO_2$ 1.435 $\mu$m bands overlap significantly.

In the south polar region, in CRISM spectra that lacked any $H_2O$ ice signatures, *Brown et al.* [2009] reported $CO_2$ ice grain sizes of up to 7 cm during southern winter. Whether large grain $CO_2$ ice deposits are present in the north during winter remains an open question.

## CONCLUSIONS

We have presented the results of CRISM and MARCI mapping of the north polar mid-winter and springtime during the first three Mars years of Mars Reconnaissance Orbiter science mapping operations.

It has been established that the optical surface of the northern polar seasonal cap is dominated by $H_2O$ ice in much the same way $CO_2$ ice dominates in the south [*Brown, et al.*, 2009].

The observations reported herein have for the first time:
1.) mapped the springtime retreat of the seasonal cap in the north polar region over MY28-30 and divided the retreat into Pre-sublimation, Early Spring, Asymmetric Retreat and Stable phases, according to the distributions of the $H_2O$ ice grain size for each period (with modes of 10±1, 30±3, 40±4 and 50±5 microns respectively).
2.) established that the apparent disappearance rate of $CO_2$ ice signatures increases rapidly in early spring and becomes spatially asymmetric after $L_s$=25°, and apparent $CO_2$ signatures disappear by $L_s$=62°.
3.) presented a model to account for the asymmetrical disappearance of $CO_2$ ice signatures during the $L_s$=25°-62° period (Figure 8).
4.) discovered only one locality where 'pure $CO_2$' ice signatures are present in mid-to-late winter in the 75°N-55°N latitude range at Lomonsov Crater at 65°N, 350°E. Regions poleward of 75°N were not observed in mid-late winter due to low light conditions.
5.) established that during late spring, 1.25 $\mu$m band depths of ~0.2 (corresponding to modeled grain sizes of 1.2 mm) are achieved in the water ice outlier regions (e.g. 150°E, 75°N).

6.) established that $H_2O$ ice grain sizes increase from mid-winter throughout springtime in contrast to the behavior of $CO_2$ ice grain size in the south pole, which decreases in size throughout springtime [*Brown, et al.*, 2009].
7.) established that the increase in grain size at 'Point B' during summer reported by *Langevin et al.* [2005] actually commences around $L_s$=86° and the grain size is remarkably stable at around 100 microns during most of spring from $L_s$=34°-86°.
8.) established that the grain size of water ice in Korolev crater increases from the time the $CO_2$ cap disappears at $L_s$=46°, most likely due to thermal metamorphism.

Digital copies of high resolution maps produced for this north polar study (and the *Brown et al.* [2009] south polar study) are available at the author's website, http://abrown.seti.org.

## ACKNOWLEDGEMENTS

We thank Ted Roush for kindly supplying his palagonite optical constants. This work would not have been possible without the outstanding efforts of the CRISM Team at JHU APL, and the staff at Malin Space Systems. NASA Grant NNX08AL09G partially funded this investigation.